\documentclass[12pt]{iopart}

\usepackage{iopams,graphicx}
\begin{document}

\title[BNS merger in Numerical Relativity]{Gravitational waves, neutrino emissions, and effects
of hyperons in binary neutron star mergers}

\author{Kenta Kiuchi, Yuichiro Sekiguchi, Koutarou Kyutoku, Masaru Shibata}

\address{Yukawa Institute for Theoretical Physics, Kyoto University, Kyoto, 606-8502, Japan}
\ead{kiuchi@yukawa.kyoto-u.ac.jp}
\begin{abstract}

Numerical simulations for the merger of binary neutron stars are performed in full general relativity incorporating both nucleonic and hyperonic finite-temperature
equations of state (EOS) and neutrino cooling. It is found that for the nucleonic and hyperonic EOS, a hyper massive neutron star (HMNS) 
with a long lifetime $(t_{\rm life}\gtrsim 10~{\rm ms})$ is the outcome for the total mass $\approx 2.7~M_\odot$. 
For the total mass $\approx 3~M_\odot$, a long-lived (short-lived with $t_{\rm life}\approx 3~{\rm ms}$) HMNS is the outcome for the nucleonic (hyperonic) EOS. 
It is shown that the typical total neutrino luminosity of the HMNS is $\sim 3$ -- $6 \times 10^{53}~{\rm erg /s}$ and the effective amplitude of gravitational waves from
the HMNS is $1$ -- $4\times 10^{-22}$ at $f\approx 2$ -- $3.2~{\rm kHz}$ for a source of distance of 100 Mpc. 
During the HMNS phase, characteristic frequencies of gravitational waves shift to a higher frequency for the hyperonic EOS in contrast to the nucleonic EOS 
in which they remain constant approximately. Our finding suggests that the effects of hyperons are well imprinted in gravitational wave and 
its detection will give us a potential opportunity to explore the composition of the neutron star matter. 
We present the neutrino luminosity curve when a black hole is formed as well.

\end{abstract}

\maketitle

\section{Introduction}
Coalescence of binary neutron stars (BNS) is drawing attention of researchers in various fields
because it is one of most promising source for next generation kilo-meter-size gravitational-wave (GW) detectors~\cite{LIGO} and
a possible candidate for the progenitor of short-hard gamma-ray bursts~\cite{SGRB} as well as a high-end laboratory of the nuclear theory~\cite{Lattimer}.

BNSs evolve due to gravitational radiation reaction and eventually merge. Before the merger sets in, each neutron star (NS) is cold, (i.e., thermal energy of constituent
nucleons is much smaller than the Fermi energy), because thermal energy inside the NSs is significantly reduced by neutrino and photon coolings due to the
long-term evolution (typically $\gtrsim~10^8$ yrs) until the merger. By contrast, after the merger, shocks are generated by hydrodynamic interactions, which
heat a merged object, hyper massive neutron star (HMNS), up to $\sim 30-50$ MeV, and hence, copious neutrinos are emitted. 
The rest mass density $\rho$ of the merged object will go well beyond the nuclear matter density $\rho_{\rm nuc}\approx 2.8\times 10^{14}{\rm g/cm^3}$ after the merger. 
In such a high density region, properties of the NS matter are still poorly understood. 
In recent years, the influence of non-nucleonic degrees of freedom, such as hyperons, meson condensations, and quarks, on properties of the NS matter has been discussed
extensively. Among exotic particles, $\Lambda$ hyperon are believed to appear first in (cold) NS around the rest mass density of $\rho\sim2-3~\rho_{\rm nuc}$
~\cite{Lattimer,Schaffner:1995th}. The presence of such the exotic particle results in softening of equation of state (EOS). Hence, if it exists or appears in the course of BNS mergers, it will 
affect the dynamics of BNS mergers and be imprinted in emitting GWs as a result. 

Motivated by these facts, numerical simulations have been extensively performed for the merger of BNS in the framework of full general relativity in the past decade
since the first success in 2000 \cite{Shibata:1999wm}. In particular, it is mandatory to do numerical relativity simulations of BNS mergers implementing a 
finite temperature EOS as well as a neutrino cooling. 
In this proceeding, we present the results based on Refs.~\cite{Sekiguchi:2011zd,Sekiguchi:2011mc}, in which the NSs are 
modeled with a nuclenoic or hyperonic EOS and the cooling is properly taken into account. 
Specifically, we adopt Shen-EOS for the nucleonic EOS \cite{Shen:1998by} and a finite-temperature EOS including contributions of $\Lambda$ hyperons as the hyperonic EOS \cite{Shen:2011qu}. 
In the following, we report a comparison between the models with these EOSs, particularly focused on the dynamics of BNS merger, emitted GWs, and neutrinos.

\section{Framework}
\subsection{Formulation and EOS}
Numerical simulations in full general relativity are performed using the following formulation and numerical schemes: Einstein's equations are solved in the 
original version of Baumgarte-Shapiro-Shibata-Nakamura formulation implementing the puncture method~\cite{BSSN}; a fourth-order finite differencing in space and a fourth-order Runge-Kutta time integration 
are used; a conservative shock capturing scheme with third-order accuracy in space and fourth-order accuracy in time is employed for solving hydrodynamic equations~\cite{Sekiguchi:2011zd,Sekiguchi:2011mc}.
We solve evolution equations for neutrino $(Y_\nu)$, electron $(Y_e)$, and total lepton $(Y_l)$ fractions per baryon, taking into account weak interaction processes 
and neutrino cooling employing a general relativistic leakage scheme for electron type $(\nu_e)$, electron anti-type $(\bar{\nu}_e)$, and other types $(\mu/\tau)$ of 
neutrinos $(\nu_x)$ \cite{Sekiguchi:2010ep,Sekiguchi:2010ja}. We implement Shen-EOS~\cite{Shen:1998by}, tabulated in terms of the rest-mass density $(\rho)$, temperature $(T)$, and $Y_e$ or $Y_l$, as the nucleonic EOS 
and a finite-temperature EOS including contributions of $\Lambda$ hyperons (Hyp-EOS) as the hyperonic EOS~\cite{Shen:2011qu}. These EOSs produce a maximum mass of zero-temperature spherical NS 
of $\approx 2.2 M_\odot$ for Shen-EOS and $\approx 1.8 M_\odot$ for Hyp-EOS. Although the maximum mass of Hyp-EOS is slightly smaller than the observational constraint of PSR J1614-2230 
$(M_{J1614-2230}=1.97 \pm 0.04 M_\odot)$~\cite{Demorest}, it deserves employing to explore the impact of hyperons on the BNS merger because it can be a viable of EOS of the NS matter 
not in extremely high densities and hence for studying the evolution of the HMNS that do not have the extremely high densities for most of their lifetime. 
It should be also noted that it is possible to produce a more massive neutron star with hyperons by fine-tuning the parameters (see e.g, in Ref.~\cite{exotic2}).

\subsection{Numerical issues and set up}
Numerical simulations are performed preparing a non-uniform grid as in Ref.~\cite{Sekiguchi:2011zd,Sekiguchi:2011mc,Kiuchi:2009jt,Kiuchi:2010ze}. 
The inner domain is composed of a finer uniform grid and 
the outer domain of a coarser nonuniform grid. The grid resolution in the inner zone is typically chosen to cover the major diameter of each NS in 
the inspiral orbit by 80 grid points. Outer boundaries are located in a local wave zone at $\approx 560-600 {\rm km}$, which is longer than 
gravitational wavelengths in the inspiral phase. 
As a monitor of the accuracy, we check the conservation of baryon rest-mass, total gravitational mass, and total angular momentum and find that 
they are preserved within the accuracy of 0.5$\%$, 1$\%$, and 3$\%$, respectively. 

References~\cite{Sekiguchi:2011zd,Sekiguchi:2011mc} focus on the merger of equal mass BNS with total mass ranging from $2.7~M_\odot$ to $3.2~M_\odot$.
In this proceeding, we report the models with total mass $\approx 2.7~M_\odot$ and $\approx 3~M_\odot$.  
There are two possible fates of BNS: If its total mass $M$ is greater than a critical mass $M_{\rm c}$, a black hole (BH) will be formed soon after the onset of 
the merger, while a differentially rotating HMNS will be formed for $M<M_{\rm c}$. $M_{\rm c}$ for Shen-EOS (Hyp-EOS) is $2.8$ -- $2.9~M_\odot$ ($2.3$ -- $2.4~M_\odot$). 
Therefore, for the binaries of $M \approx 2.7~M_\odot$, we expect that Shen-EOS model will form a HMNS and Hyp-EOS model will collapse to a BH. 
The models with $M \approx 3~M_\odot$ are anticipated to collapse to the BH, irrespective of the EOSs. 
We name Shen-EOS (Hyp-EOS) model with $M \approx 2.7~M_\odot$ S135 (H135) and those with $M \approx 3~M_\odot$ S15 (H15). 
It should be noted that these figures of the mass are motivated by the narrow mass distribution of observed binary neutron stars.

\section{Numerical results}
\begin{figure}
\begin{center}
\includegraphics[scale=1.0]{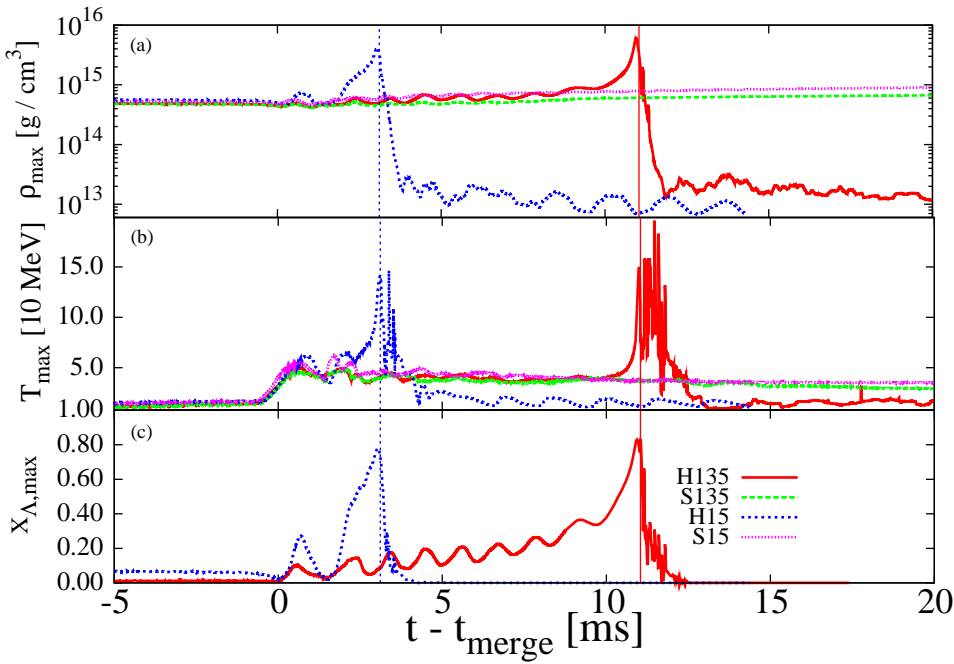}
 \caption{Maximum rest-mass density, maximum matter temperature, and
   maximum mass fraction of hyperons as functions of time for H135 (solid-red), S135 (dashed-green), 
   H15 (short-dashed blue), and S15 (dotted-magenta). 
   The vertical thin solid (short-dashed) line shows the time at which a BH is formed for H135 (H15).
   \label{fig1}}
\end{center}
\end{figure}

Figure \ref{fig1} plots the evolution of
maximum rest-mass density, $\rho_{\rm max}$, maximum temperature,
$T_{\rm max}$, and maximum mass fraction of hyperons, $X_{\Lambda, {\rm max}}$
as functions of $t-t_{\rm merge}$ where $t_{\rm merge}$ is an approximate 
onset time of the merger. Before the merger ($t<t_{\rm
merge}$), $\rho_{\rm max}$ and $T_{\rm max}$ for H135 (H15) agree
well with those for the corresponding S135 (S15), because $X_{\Lambda, {\rm max}}$ in this phase is small as
$O(10^{-2})$ and effects of hyperons on dynamics are not significant. 
After the merger sets in ($t>t_{\rm merge}$), on the other hand,
$X_{\Lambda,{\rm max}}$ increases to be $\gtrsim 0.1$ in accordance with the
increase in $\rho_{\rm max}$, and hyperons play a substantial role in 
the post-merger dynamics. 
The panel (b) in Fig.~\ref{fig1} tells us that $T_{\rm max}$ in the HMNSs reach to $50$ -- $60$ MeV just after the first 
contact, irrespective of the models. For H15, it rapidly increases to a high value of $130$ -- $140$ MeV just before the collapse to a BH.
For H135 and Shen-EOS models, it decreases due to the neutrino cooling in the subsequent evolution and then relaxes 
to $\sim 25$ -- $40$ MeV. We observe the rapid enhancement of $T_{\rm max}$ just before a BH formation in H135. 

Although the total mass, $M$, is larger than the maximum mass of
the zero-temperature spherical NS for all the models, a HMNS is formed
after the merger, supported by the centrifugal force and thermal
contribution to the pressure~\cite{Sekiguchi:2011zd,Hotokezaka:2011dh}. 
In particular, the enhancement of the critical mass $M_{\rm c}$ due to the thermal effect is prominent in S15, which 
does not collapse to a BH during the simulation contrary to our expectation. 
The HMNSs subsequently contract by emission of GWs, which carry energy and angular momentum from the HMNS; $\rho_{\rm max}$ 
increases in the gravitational radiation timescale. 
For Hyp-EOS models, they collapse to the BH at $t=t_{\rm BH}$ where
$t_{\rm BH}-t_{\rm merge} \approx 11.0$ ms for H135 and $\approx 3.1$ ms for H15. 
After the BH formation of Hyp-EOS models, an accretion torus is formed around the BH and gradually settles down to a 
quasi-steady state. The torus mass is $\approx$ 0.082, and 0.035 $M_\odot$ for H135 and H15, respectively~\cite{Sekiguchi:2011mc}. 
The torus mass of H135 is greater than that of H15 due to the longer transient HMNS phase, in which the angular momentum is transported outwardly.
Note that the lifetime of HMNS for H135 is longer than that for H15. 
At $t-t_{\rm merge}\sim 20~{\rm ms}$ for Shen-EOS models, the degree of its nonaxial symmetry becomes low enough that 
the emissivity of GWs is significantly reduced. Because no dissipation process except for the neutrino cooling is present, 
the HMNS has not yet collapsed to a BH. We expect it will be alive at least for a cooling time, $t_{\rm cool} \equiv E_{\rm th}/L_{\nu} \sim 2$--3 s,
where $E_{\rm th}$ is total thermal energy and $L_{\nu}$ is total
neutrino luminosity~\cite{Sekiguchi:2011zd}.

\begin{figure}
\begin{center}
\includegraphics[scale=1.0]{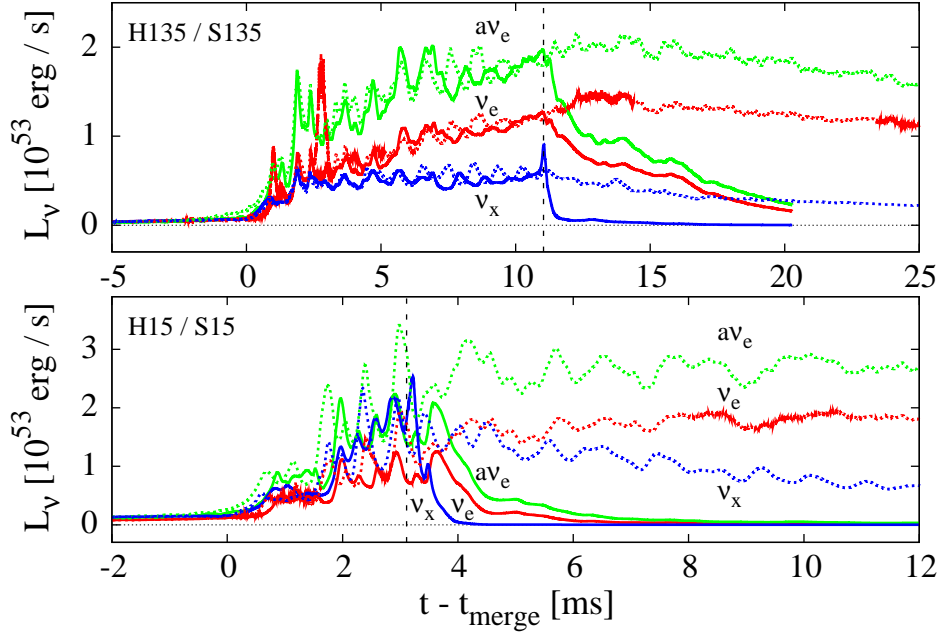}
 \caption{Neutrino luminosities for H135, S135, H15 and S15. The solid (short-dashed) curves 
   are for Hyp-EOS (Shen-EOS) models. The dashed vertical lines show
   the time at which a BH is formed for Hyp-EOS models and $a{\nu}_e$ represents electron anti neutrino.
   \label{fig2}}
\end{center}
\end{figure}

Figure \ref{fig2} plots neutrino luminosities as functions of time for 
three flavors ($\nu_e,\bar{\nu}_e$, and sum of $\nu_x$). It is found that 
electron anti neutrinos are dominantly emitted for any model. The reason for this is as follows: 
The HMNS has a high temperature, and hence, electron-positron pairs are efficiently produced from thermal photons, in particular 
in its envelope. 
The positron capture $n+e^+\to p + \bar{\nu}_e$ proceeds more preferentially than the electron capture 
$p+e^-\to n +\nu_e$ because the proton fractions is much smaller than the neutron fraction. 
Such hierarchy in the neutrino luminosities was reported also in 
Ref.~\cite{Ruffert}. 
These features in neutrino luminosities are quantitatively the same for Hyp-EOS and
Shen-EOS models, and hence, it would be difficult to extract
information of the NS matter only from the neutrino signal. 
Soon after the BH formation for Hyp-EOS models, $\mu/\tau$ neutrino luminosity steeply decreases because
high temperature regions are swallowed into the BH, while luminosities
of electron neutrinos and anti neutrinos decrease only gradually
because these neutrinos are emitted via charged-current processes from
the massive accretion torus.  

The anti neutrino luminosity for the long-lived HMNS for Shen-EOS models is $L_{\rm \bar{\nu}}\sim 2$ -- $3\times 10^{53}{\rm erg/s}$ with small time variability. 
It is greater than that from the protoneutron stars found after supernovae \cite{Sumiyoshi:2005ri}. Averaged neutrino energy is $\epsilon_{\bar{\nu}}\sim 20-30~{\rm MeV}$. 
The sensitivity of water-Cherenkov neutrino detectors such as Super-Kaminokande and future Hyper-Kamiokande (HK) have a good sensitivity for such high-energy neutrinos in particular for 
electron anti neutrinos. The detection number for electron neutrinos is approximately estimated by $\sigma \Delta T L_{\bar{\nu}}/(4\pi D^2 \epsilon_{\bar{\nu}})$ where $\sigma$ is the total 
cross section of the detector against target neutrinos, $\Delta T$ is the lifetime of the HMNS, and D is the distance to the HMNS. For one-Mton detector such as HK, the expected 
detection number is $\gtrsim 10$ for $D \lesssim 5$ Mpc with $\Delta T\sim 2$ -- 3 s, 
neutrinos from the HMNS may be detected and its formation may be confirmed. Note that GWs from the HMNS will be simultaneously
detected for such a close event, reinforcing the confirmation of the HMNS formation. 
For Hyp-EOS models, the detection of anti neutrino is less likely because of the short lifetime of the HMNSs. However, as discussed below, 
the gravitational waves could constrain the hyperonic or nucleonic EOS. 

\begin{figure}
\begin{center}
  \vspace*{40pt}
    \begin{tabular}{cc}
      \begin{minipage}{0.5\hsize}
      \includegraphics[scale=0.72]{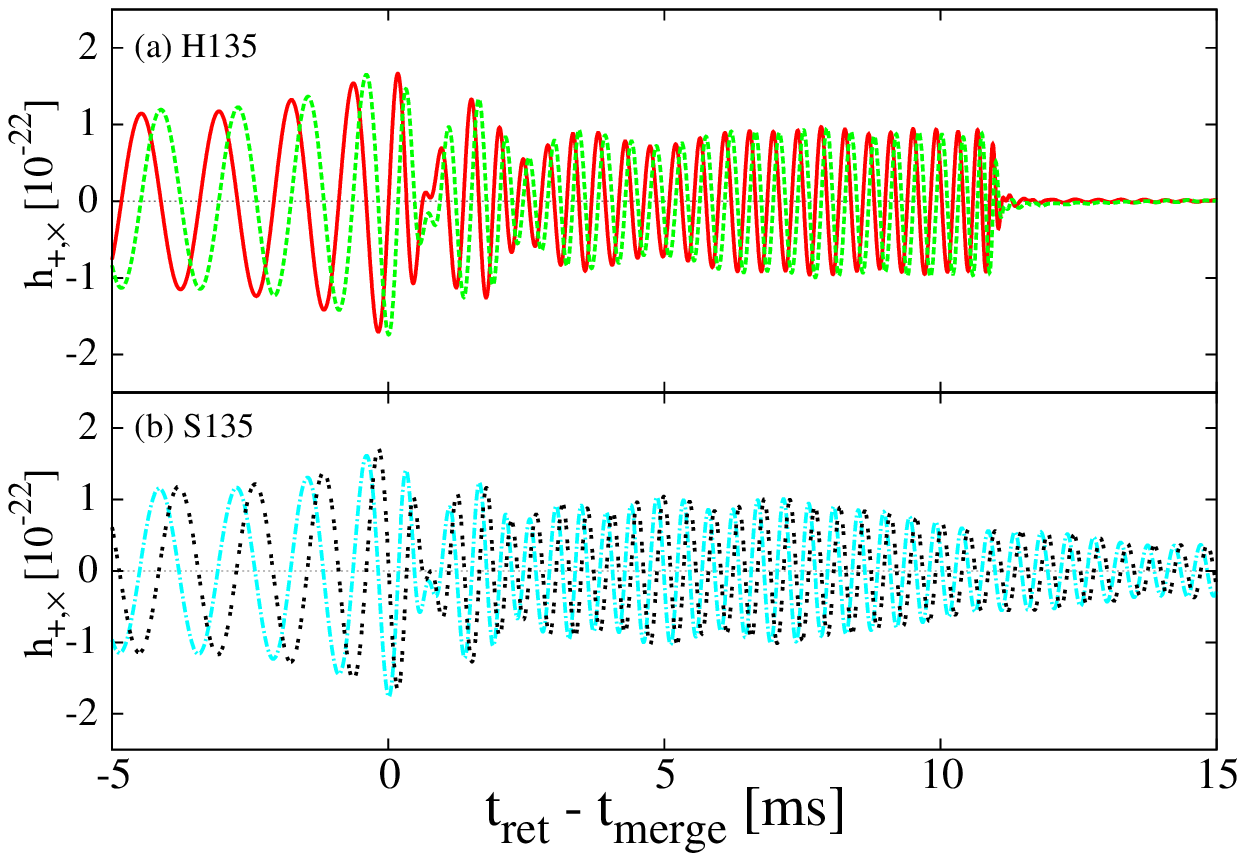}
      \end{minipage}
      \hspace{0.9cm}
      \begin{minipage}{0.5\hsize}
      \includegraphics[scale=0.72]{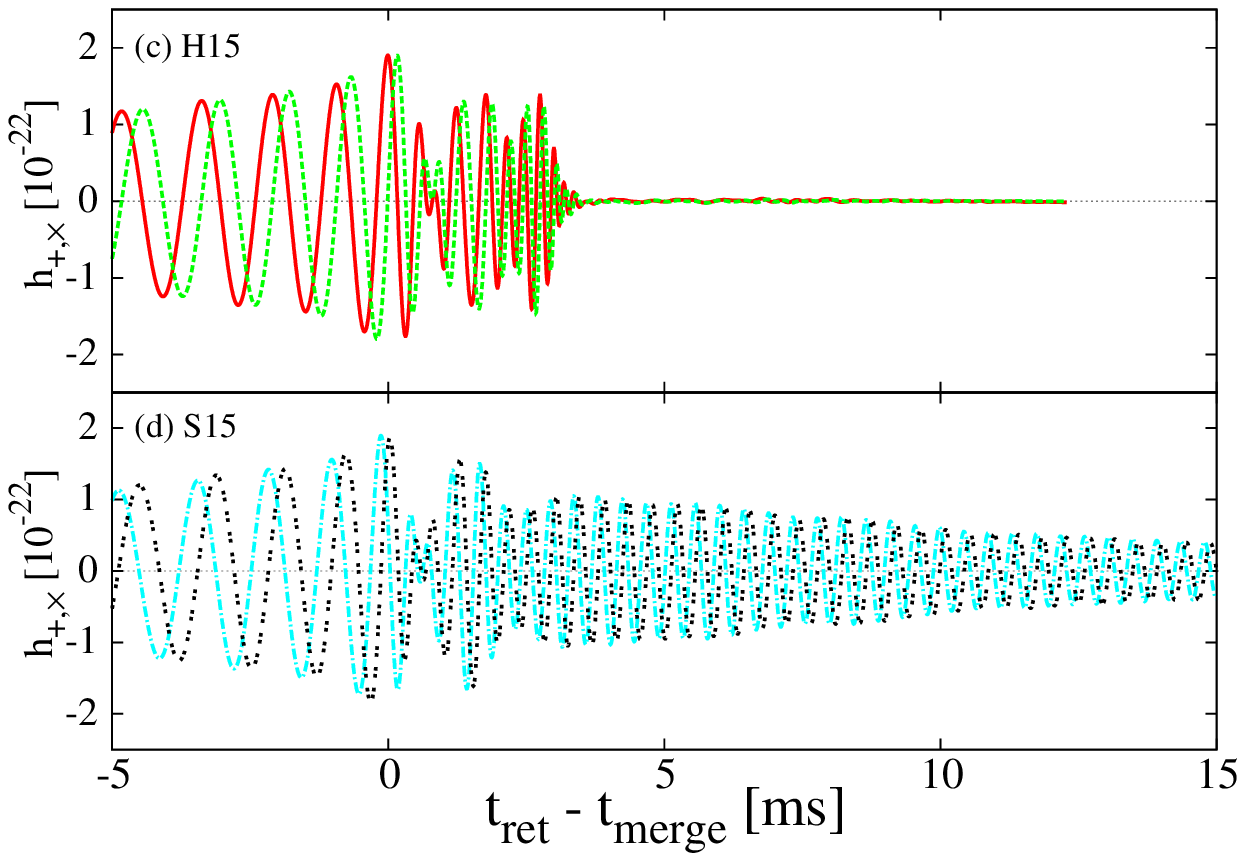}
      \end{minipage}
    \end{tabular}
\caption{GWs observed along the
axis perpendicular to the orbital plane for the hypothetical distance
to the source $D=100$~Mpc for (a) H135, (b) S135, (c) H15, and (d) S15.
   \label{fig3}}
\end{center}
\end{figure}

\begin{figure}
\begin{center}
  \vspace*{40pt}
    \begin{tabular}{cc}
      \begin{minipage}{0.5\hsize}
      \includegraphics[scale=0.72]{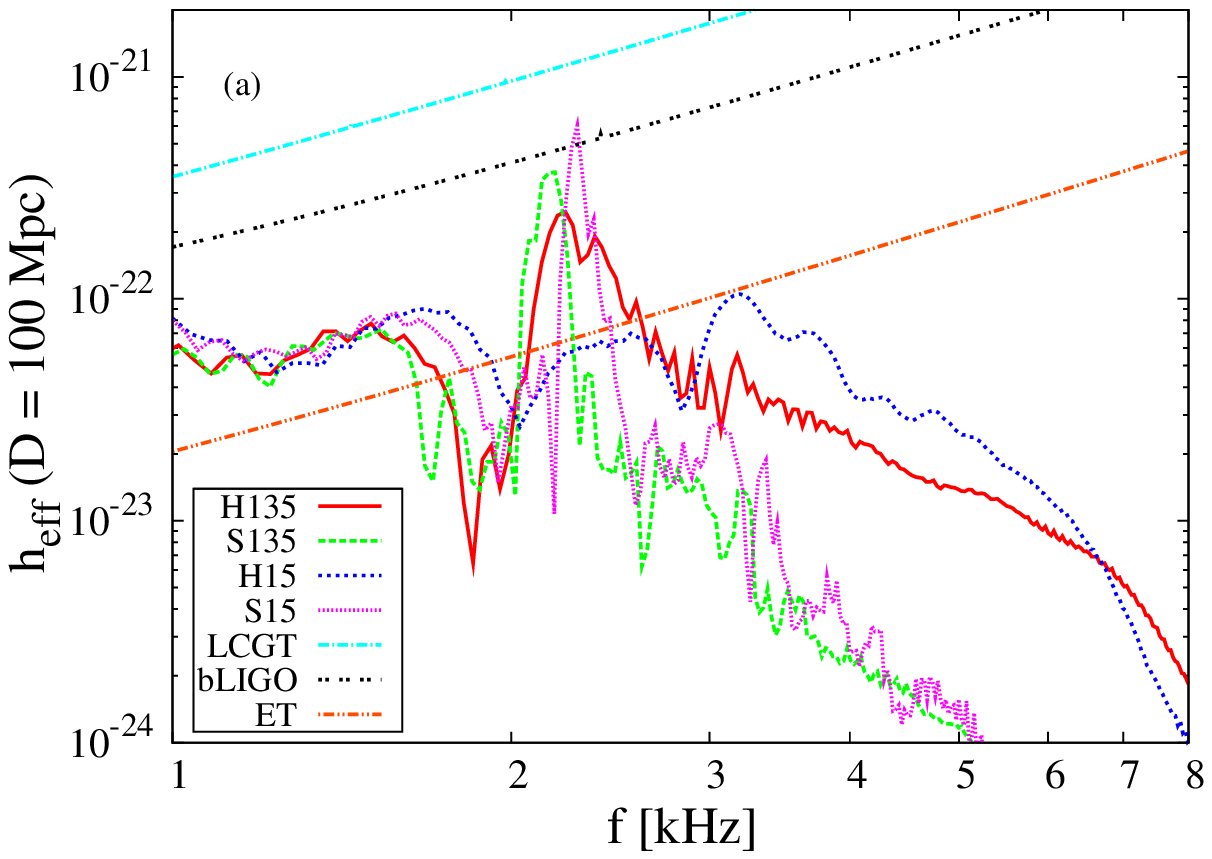}
      \end{minipage}
      \hspace{0.9cm}
      \begin{minipage}{0.5\hsize}
      \includegraphics[scale=0.72]{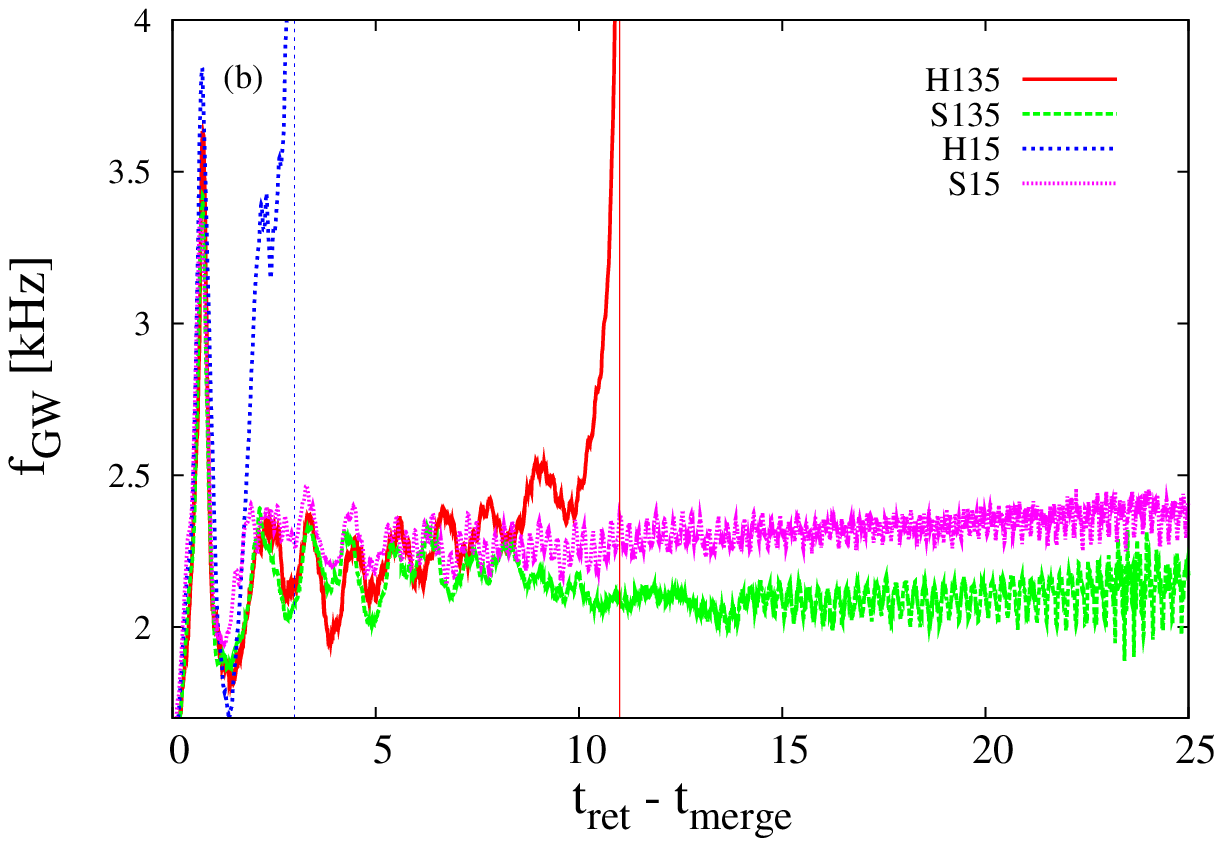}
      \end{minipage}
    \end{tabular}
\caption{
(a) The effective amplitude of
GWs defined by $0.4f|h(f)|$ as a function of frequency for $D=100$~Mpc.
The noise amplitudes of a broadband configuration of
Advanced Laser Interferometer Gravitational wave Observatories (bLIGO), 
Large-scale Cryogenic Gravitational wave Telescope (LCGT) and 
Einstein Telescope (ET) are shown together. 
(b) The frequency of gravitational wave in the HMNS phase for H135 (solid-red), S135 (dashed-green),
   H15 (short-dashed blue), and S15 (dotted-magenta). The thin vertical lines are BH formation time 
   for Hyp-EOS models.
   \label{fig4}}
\end{center}
\end{figure}

Figure \ref{fig3} plots the plus and cross mode ($h_{+,\times}$) of GWs with 
$l=|m|=2$ as a function of $t_{\rm ret}-t_{\rm merge}$ where $t_{\rm ret}$ is the
retarded time $t_{\rm ret}=t-D-2M\log(D/M)$,
extracted from the metric in the local wave zone, i.e., at $\approx$ 550 km. 
The waveforms are composed of the so-called chirp waveform, which is emitted when
the BNS is in an inspiral motion (for $t_{\rm ret}\lesssim t_{\rm merge}$), and the merger
waveform (for $t_{\rm ret}\gtrsim t_{\rm merge}$).
The GW amplitude is $|h_{+,\times}| \lesssim 2\times 10^{-22}$ for a source at a
distance $D=100$ Mpc with the direction perpendicular to the orbital plane. GWs from the inspiral phase (for $t_{\rm ret}
\lesssim t_{\rm merge}$) agree well with each other for the models
with Hyp-EOS and Shen-EOS with the same mass. On the other hand,
quasi-periodic GWs from the HMNS (for $t_{\rm ret} \gtrsim t_{\rm
merge}$) show differences. 
Quasi-periodic GWs is suddenly shut down at the BH formation for Hyp-EOS models, $t_{\rm ret}-t_{\rm merge} \approx 11$ ms for H135 
and $t_{\rm ret}-t_{\rm merge}\approx 3$ ms for H15.
This is because the HMNS collapses to a BH due to the softening of the EOS
before relaxing to a stationary spheroid. The quasi normal mode is excited after the BH formation, 
which is imprinted in the gravitational wave spectra at $\sim 7$ kHz (see Fig.~\ref{fig4}(b)). 
For Shen-EOS models, the long-lived HMNSs emit the gravitational waves whose amplitude 
gradually decreases indicating that they approach to the axi-symmetric quasi-stationary state. 

The characteristic GW frequency, $f_{\rm GW}$, increases with
time for Hyp-EOS models. 
These facts are well imprinted in the effective amplitude (see Fig.~\ref{fig4}(a))
defined by $h_{\rm eff}
\equiv 0.4 f |h(f)|$ where $h(f)$ is the Fourier transform of $h_+ -i
h_{\times}$ and the factor 0.4
comes from taking average in terms of direction to the source and
rotational axis of the HMNS. 
The peak amplitudes of $h_{\rm eff}(f)$ in Hyp-EOS models are 
smaller than those in Shen-EOS models due to a shorter lifetime of the HMNS. 
The prominent peak in the GW spectrum for Hyp-EOS 
models is broadened because of the shift of the characteristic frequency. 
This frequency shift is well imprinted in the evolution of the characteristic frequency 
$f_{\rm GW}$ in Fig.~\ref{fig4} (b), where $f_{\rm GW} \equiv d\phi_{\rm GW}/dt/2\pi$ with the 
Weyl scalar $\Psi_4=A {\rm e}^{-i \phi_{\rm GW}}$. For Shen-EOS models, the central value of the frequency remains almost constant 
after the merger $(t_{\rm ret}\gtrsim t_{\rm merge})$ and it corresponds to the peak frequency of gravitational wave spectrum in 
Fig.~\ref{fig4} (a). By contrast, Hyp-EOS models exhibit a gradual increase in $f_{\rm GW}$. Specifically, 
for H135, $f_{\rm GW}$ grows from $\approx 2 $ kHz at $t_{\rm ret}-t_{\rm merge}=2$ ms to $\approx 2.5$ kHz at $t_{\rm ret}-t_{\rm merge}=10$ ms.
For H15, it increases from $\approx 2 $ kHz at $t_{\rm ret}-t_{\rm merge}=2$ to $\approx 3.4$ kHz at $t_{\rm ret}-t_{\rm merge}=2.5$ ms though the lifetime of HMNS is short. 
This feature in the frequency shift is explained as follows. For Shen-EOS models, the gravitational waves carry away the angular momentum and then  
the HMNS slightly contracts. For Hyp-EOS models, the hyperon fraction $X_{\Lambda}$ increases with the contraction of the HMNS, 
resulting in the relative reduction of the pressure. This leads further contraction of the HMNS. 
Recent studies suggest that $f_{\rm GW}$ is associated with the f-mode which is approximately proportional to 
$\sqrt{M_{\rm NS}/R_{\rm NS}^3}$ with $M_{\rm NS}$ and $R_{\rm NS}$ being the mass and radius of HMNS~\cite{Stergioulas:2011gd,Bauswein:2011tp}. 
Applying this result to our case indicates that Hyp-EOS models exhibit the frequency shift due to the run-away contraction, i.e., 
decreasing the radius of the HMNSs, and 
the Shen-EOS models emit the gravitational waves with the coherent frequency due to the slight contraction. 
In other words, the frequency $f_{\rm GW}$ for Shen-EOS models evolves on a timescale of gravitational radiation reaction. 
The effective amplitude $1$ -- $4\times 10^{-22}$ for $D=100$ Mpc suggests that
for a specially-designed version of advanced GW detectors such as broadband LIGO, which has a
good sensitivity for a high-frequency band, GWs from the HMNS oscillations may be detected with ${\rm S}/{\rm N}=5$
if $D\lesssim 20$ Mpc or the source is located in an optimistic direction~\cite{Sekiguchi:2011zd}.

\section{Summary}
We have reported the results of numerical-relativity simulation performed incorporating both a finite-temperature nucleonic EOS and
hyperonic EOS as well as neutrino cooling effect. We showed that for both the nucleonic and hyperonic EOS, HMNS is the canonical outcome and BH is not promptly
formed after the onset of the merger for the canonical value of the NS mass $1.35~M_\odot$ and for more massive case of $1.5M_\odot$. 
The primary reason is that the thermal pressure plays an important role for sustaining the HMNS. 
For the nucleonic EOS, the lifetime of the formed HMNS is much longer than its dynamical timescale, $\gg 10{\rm ms}$, and will be determined by the timescale of neutrino
cooling. For the hyperonic EOS, the HMNS subsequently collapse to a BH due to a softening of EOS as a consequence of increase of $\Lambda$ hyperon.
Neutrino luminosity of the HMNS was shown to be high as $\sim 3$ -- $6\times 10^{53} {\rm erg/s}$. 
The effective amplitude of GWs is $1$ -- $4\times 10^{-22}$ at $f_{\rm peak}\approx 2$ -- $3.2~{\rm kHz}$
for a source distance of 100 Mpc. 

We further found that the characteristic frequency of gravitational waves, $f_{\rm GW}$, from the HMNS increases with time for the hyperonic EOS in contrast to the 
nucleonic EOS in which $f_{\rm GW}$ approximately remains constant. 
This time-dependent characteristic frequency leads to the broaden gravitational wave spectrum for the hyperonic EOS. 
For the nucleonic EOS, the spectrum has a sharp peak around the less time-dependent frequency. 
This result suggests that the emergence of hyperons may be captured from 
the evolution of characteristic frequency and the peak width of the gravitational waves spectra. 
If the BNS merger happens at a relatively short distance or is located in an optimistic direction, such GWs may be detected and HMNS formation will be
confirmed. In particular, to increase the chance of the detection of such GWs,
we need for the detectors with high sensitivities at high frequencies of several kilohertz such as the Einstein Telescope~\cite{ET}.

\small\normalsize

\small\normalsize

\small\normalsize

\small\normalsize

\ack Numerical simulations are performed on SR16000 at YITP of Kyoto University and SX9 and XT4 at CfCA of NAOJ.
This work was supported by Grant-in-Aid for Scientific Research (21018008, 21105511, 21340051, 22740178, 23740160), Grant-in-Aid on
Innovative Area (20105004), and HPCI Stragetic Program in Japan MEXT.

\end{document}